\documentclass[aps,prb,10pt,twocolumn,showpacs,superscriptaddress]{revtex4-1}
\usepackage{amsmath, amssymb, amsfonts}
\usepackage{graphicx}
\usepackage{bm}
\usepackage[squaren]{SIunits}

\providecommand{\vect}[1]{{\boldsymbol{#1}}}

\usepackage{color}

\begin{document}

\title{Current-driven periodic domain wall creation in ferromagnetic nano-wires}

\author{Matthias Sitte}
\affiliation{Institute of Physics, Johannes Gutenberg-Universit{\"a}t, 55128 Mainz, Germany}

\author{Karin Everschor-Sitte}
\affiliation{Institute of Physics, Johannes Gutenberg-Universit{\"a}t, 55128 Mainz, Germany}

\author{Thierry Valet}
\affiliation{Institute of Physics, Johannes Gutenberg-Universit{\"a}t, 55128 Mainz, Germany}

\author{Davi~R. Rodrigues}
\affiliation{Department of Physics \& Astronomy, Texas A\&M University, College Station, Texas 77843-4242, USA}

\author{Jairo Sinova}
\affiliation{Institute of Physics, Johannes Gutenberg-Universit{\"a}t, 55128 Mainz, Germany}

\author{Ar.~Abanov}
\affiliation{Department of Physics \& Astronomy, Texas A\&M University, College Station, Texas 77843-4242, USA}

\date{\today}

\begin{abstract}
We predict the electrical generation and injection of domain walls into a ferromagnetic nano-wire without the need of an assisting magnetic field.  Our analytical and numerical results show that above a critical current $j_{c}$ domain walls are injected into the nano-wire with a period $T \sim (j-j_{c})^{-1/2}$.  Importantly, domain walls can be produced periodically even in a simple exchange ferromagnet with uniaxial anisotropy, without requiring any standard ``twisting'' interaction like Dzyaloshinskii-Moriya or dipole-dipole interactions.  We show analytically that this process and the period exponents are universal and do not depend on the peculiarities of the microscopic Hamiltonian.  Finally we give a specific proposal for an experimental realization.
\end{abstract}

\pacs{}

\maketitle

Recent proposals for the next generation of magnetic memory devices\cite{Parkin2008, Jiang2010, Tomasello2014} rely on the ability to manipulate the position and orientation of domain walls (DW) in ferromagnetic nano-wires by electric current.\cite{Tatara2004, Zhang2004, Thiaville2005, Brataas2012, Tretiakov2012}  These proposals have led to an intense research activity in the area of current-induced DW dynamics in both anti- and ferromagnets.\cite{Swaving2011, Hals2011, Tveten2014, Shiino2016, Gomonay2016}  However, to manipulate DWs one needs to first create them.  Currently, DWs are injected from one end into a nano-wire by applying a magnetic field.  In this work we propose a technique to controllably and reliably inject the DWs into nano-wires by just applying a DC electric current without the need of magnetic fields.  Our finding leads to the possibility to make an ``all electric'' DW-dynamics-based spintronic devices.

\begin{figure}
\includegraphics[width=0.95\columnwidth]{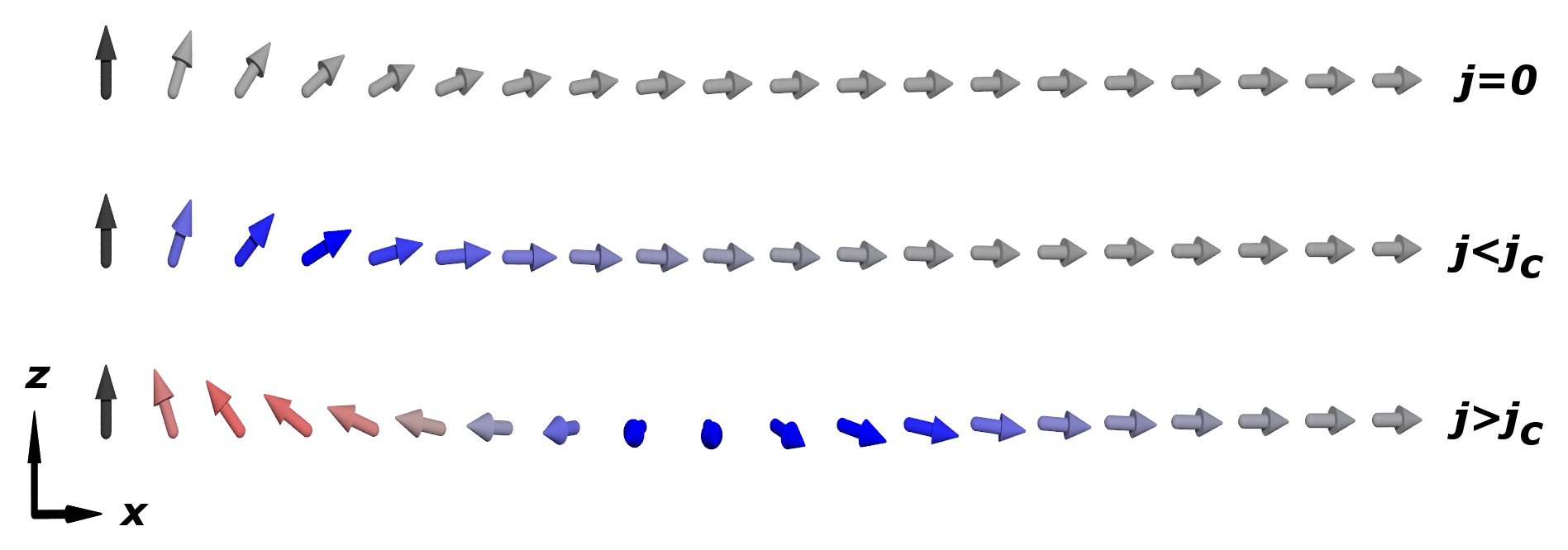}
\caption{(Color online) Magnetization texture for increasing currents.  Without current ($j = 0$, top) the configuration is half of a planar DW centered at $x=0$; for $j < j_{c}$ the domain wall is tilted out of the $xz$ plane; for currents larger than $j_{c}$ (bottom panel) domains move along the nano-wire.  The black arrow at the start of the wire is fixed, e.g., by an adjacent permanent ferromagnet (not shown).  The color code is chosen such that gray arrows lie in the $xz$ plane, while red (blue) arrows have a finite component out of the $xz$ plane in $\pm y$ direction, respectively.}
\label{fig:fig1}
\end{figure}

Within this paper we consider a magnetic nano-wire where at one end the magnetization is fixed (e.g., by an adjacent permanent ferromagnet) along a different direction than the anisotropy direction of the wire (see Fig.~\ref{fig:fig1}).  Close to the fixed end, the magnetization will twist naturally on a length scale defined by the interplay of the fixed magnetization and the anisotropy strength and in the plane defined by the direction of the fixed magnetization and the anisotropy direction.  We show analytically and numerically that for DC currents larger than a critical current $j_c$ even such a simple structure becomes unstable, and DWs are produced periodically with a period $T \sim (j-j_{c})^{-1/2}$.

The instability and the phenomena of periodic, current-induced DW production can be easily explained (see Fig.~\ref{fig:fig1}):  When ramping up the current strength, the current will i) twist the magnetic structure around the anisotropy direction and ii) elongate the texture close the pinned end.  For currents $j > j_{c}$ the magnetic structure is ``twisted off'' and produces a DW which then moves along the wire.

We demonstrate that under broad conditions this effect does not depend on the details of the microscopic magnetic Hamiltonian and does not require any ``twisting'' terms in the Hamiltonian such as Dzyaloshinskii-Moriya (DMI) or dipole-dipole interactions.  We compute the critical current analytically and confirm the results numerically for a specific microscopic Hamiltonian.  In the limit of a current strength just above the critical value, $j \gtrsim j_{c}$, the dynamics of the DWs is very slow and dissipation only plays a minor role.

Below we first describe the setup for the magnetic nano-wire and review the standard Landau-Lifshitz Gilbert equation to describe magnetization dynamics.  Next we show analytically and via simulations that a pinning center in a magnetic nanowire constitutes a dynamic instability in the system for currents above a critical one.  We determine the critical current density analytically and numerically and our numerical and analytical results agree within a few percent.  By mapping the magnetic problem to an effective one dimensional model of a particle in a potential, we can easily interpret the magnetic texture for currents below the critical one as  part of a DW.  This analysis also guides us to the ferromagnetic instability, where above an even larger current the ferromagnetic solution becomes unstable.  In Sec.~\ref{sec:DynamicCreationOfDWs} we discuss the dynamics of the DW creation.  Finally, we discuss the problem of minimization of Ohmic losses in the injection process as well as the relevance of the presented calculation for the dynamics around a strong pinning center relevant to experiments.

\section{Micromagnetic model and magnetization dynamics}
\label{sec:Model}

We emphasize that our results do not depend on the specific details of the micromagnetic model.  The only requirements are exchange interactions and uniaxial anisotropy along a different direction than the pinned magnetization at the one end of the wire.  Note that uniaxial anisotropy must be present in order to stabilize a DW in a nano-wire.

However, to be specific and to explicitly show the essential part of the analytic calculation, we propose for the current-induced injection of DWs into the ferromagnetic nano-wire the simple geometry as shown in Fig.~\ref{fig:fig1}:  We assume that the nano-wire is thin enough so that the magnetic configuration is one-dimensional.  The easy axis is directed along the wire which is taken to be semi-infinite going from $x=0$ to $x=\infty$.  At the start of the wire the magnetization is fixed along the $z$ direction, $\vect{M}(0) = \vect{e}_{z}$ (e.g., by an adjacent ferromagnet with a large uniaxial anisotropy along the $z$ direction).

Therefore, the free energy is given by
\begin{equation}
\label{eq:FreeEnergy}
F[\vect{M}] = \int_{0}^{\infty} \biggl[ \frac{J}{2}(\partial_{x} \vect{M})^{2} + \lambda \Pi (M_{x}) \biggr] dx,
\end{equation}
where $\vect{M}$ is the local magnetization direction with the (local) condition $\vect{M}^2(x) = 1$.  The first term is the exchange term with exchange constant $J$ and the second one describes the anisotropy term with strengh $\lambda$.  The uniaxial magnetic anisotropy along the nano-wire is typically described by the term $\lambda (1 - M_{x}^{2})$, where $\lambda$ is the anisotropy strength, and the constant term is chosen such that the free energy density of the nano-wire at infinity is zero.  In Eq.~\eqref{eq:FreeEnergy} we use a more general form $\lambda \Pi (M_{x})$, where $\Pi$ is a monotonic function with $\Pi(0)=1$, $\Pi '(0)=0$ and $\Pi(1)=0$. The boundary conditions are summarized in Table~\ref{tab:BoundaryConditions}.  To compare with numerics note that $J$ and $\lambda$ are the exchange constant and anisotropy strength per unit length, respectively, and one has to devide them by the cross section area $\sigma$ of the wire to obtain the bulk parameters: $J_{\mathrm{bulk}}= J/\sigma$ and $\lambda_{\mathrm{bulk}}= \lambda/\sigma$.

\begingroup
\squeezetable
\begin{table}[b]
\begin{ruledtabular}
\begin{tabular}{l|c c c}
& $M_{x}$ & $M_{z}$ & $\Pi(M_{x})$ \\
\hline
$x=0$ & $0$ & $1$ & $1$ \\
$x=\infty$ & $1$ & $0$ & $0$ \\
\end{tabular}
\end{ruledtabular}
\caption{Boundary conditions of magnetization components $M_{x}$ and $M_{z}$, and of the uniaxial anisotropy $\Pi$ for the chosen nano-wire geometry.}
\label{tab:BoundaryConditions}
\end{table}
\endgroup

To describe the current-induced magnetization dynamics we use the standard Landau-Lifshitz-Gilbert equation with current:
\begin{equation*}
(\partial_{t} + v_{s} \partial_x) \vect{M} = -\gamma \vect{M} \times \vect{H}_{\mathrm{eff}} + \frac{\alpha}{M_{s}} \vect{M} \times (\partial_{t} + \frac{\beta}{\alpha} v_{s} \partial_{x}) \vect{M},
\end{equation*}
where $\gamma$ is the gyromagnetic ratio, $M_{s}$ is the saturation magnetization per unit length with $M_{s}^{\mathrm{bulk}} = M_{s}/\sigma$, and $\alpha$ and $\beta$ are the adiabatic and non-adiabatic damping parameters.  The effective magnetic field is given by $\vect{H}_{\mathrm{eff}}= -M_{s}^{-1} (\delta F[\vect{M}]/\delta \vect{M})$, and the applied current along the $x$ direction enters the equation via the effective spin velocity:
\begin{equation}
v_{s} = \frac{P \mu_{B}}{e M_{s} [1 + (\beta/\alpha)^2]} j
\end{equation}
where $P$ is the current polarization, $\mu_{B}$ is the Bohr magneton, $e$ is the electron charge, and $j$ is the current with the corresponding current density $j/\sigma$.  For simplicity we assume in the following $\beta = 0$.  However we checked numerically that a finite $\beta$ does not change qualitatively the results.

Next we discuss the instability and at which critical current it arrises, followed by detailed analytical and numerical calculations of the static case below the critical current and the dynamic case above the critical current.

\section{Instability and critical current}
\label{sec:Instability}

In this section we show that for currents $j<j_c$ there is a static configuration which becomes unstable at a specific critical current $j_c$ computed below.  Since for a current density slightly larger than the critical current density $j_{c}$, the magnetization dynamics is a slow and quasi-adiabatic process, we can simply ignore dissipative terms to determine $j_{c}$ analytically.  Therefore, we can use the simplified equation
\begin{equation}
\label{eq:SimplifiedLLG}
\partial_{t} \vect{M} = -\gamma \vect{M} \times \vect{H}_{\mathrm{eff}} - v_{s} \partial_{x} \vect{M},
\end{equation}
which can be rewritten in the following form,
\begin{equation}
\label{eq:EffectiveFreeEnergy}
\partial_{t} \vect{M} = \frac{\gamma}{M_{s}} \vect{M}\times \frac{\delta F_{\mathrm{eff}}[\vect{M}, v_{s}]}{\delta \vect{M}},
\end{equation}
where
\begin{equation*}
F_{\mathrm{eff}}[\vect{M}, v_{s}] = F[\vect{M}]  + \int_{0}^{\infty} \Omega\ dx
\end{equation*}
with
\begin{equation*}
\frac{\gamma}{M_{s}} \vect{M} \times \frac{\delta}{\delta \vect{M}} \biggl( \int_{0}^{\infty} \Omega\ dx \biggr) \equiv - v_{s} \partial_{x} \vect{M}.
\end{equation*}
The Berry phase-like term $\Omega$ can be written in terms of the $\mathrm{CP}^{1}$ representation of the unit vector field $\vect{M}(x)$, but its exact form is not required here.

Eq.~\eqref{eq:EffectiveFreeEnergy} shows that any finite dissipation will ensure that the physical\footnote{Eq.~\eqref{eq:SimplifiedLLG} is non-linear and thus has many solutions satisfying the boundary conditions.} static solution minimizes $F_{\mathrm{eff}}$.  Let us now change perspective and consider the effective free energy $F_{\mathrm{eff}}$ as an action of some model and the coordinate $x$ as time.  The corresponding Lagrangian of this model is given by
\begin{equation}
\label{eq:Lagrangian}
\mathcal{L} = \frac{J}{2}(\partial_{x} \vect{M})^{2} + \lambda \Pi (M_{x}) + \Omega.
\end{equation}
The translational invariance of $F_{\mathrm{eff}}$ --- the analogue of the fact that effective free energy $F_{\mathrm{eff}}$ does not explicitly depend on ``time'' $x$ --- implies that the Hamiltonian corresponding to the Lagrangian~\eqref{eq:Lagrangian} is conserved.  This Hamiltonian is given by:\footnote{The Berry term, as usual, does not contribute to the Hamiltonian}
\begin{equation}
\mathcal{H} = \frac{\partial \mathcal{L}}{\partial (\partial_{x} \vect{M})} \cdot \partial_{x}\vect{M} - \mathcal{L} = \frac{J}{2}(\partial_{x} \vect{M})^{2} - \lambda \Pi(M_{x}).
\end{equation}
On a physical solution this Hamiltonian is conserved, i.e., it does not depend on ``time'' $x$.  Because at $x \to \infty$ the magnetization is assumed to be parallel to $\hat{\vect{x}}$ we find that the Hamiltonian has to vanish everywhere due to translational invariance, $\mathcal{H} \equiv 0$.  Consequently, the static configuration must satisfy the following relation for all $x$:
\begin{equation}
\label{eq:StaticCaseEnergy}
\frac{J}{2}(\partial_{x} \vect{M})^{2} - \lambda \Pi(M_{x}) \equiv 0.
\end{equation}
Furthermore, note that the $x$ component of the total angular momentum is conserved.  This can be derived explicitly by multiplying Eq.~\eqref{eq:SimplifiedLLG} by the unit vector in the $x$ direction:
\begin{align*}
\partial_{t} M_{x}
&= - \frac{\gamma}{M_{s}} J \hat{\vect{x}} \cdot (\vect{M} \times \partial_{x}^{2} \vect{M}) - v_{s} \partial_{x} M_{x} \\
&= - \partial_{x} \biggl[ \frac{\gamma}{M_{s}} J \hat{\vect{x}} \cdot (\vect{M} \times \partial_{x} \vect{M}) + v_{s} M_{x} \biggr].
\end{align*}
In the static case, $\partial_{t} M_{x} \equiv 0$, we can compare the right hand side with its value at infinity to obtain
\begin{equation}
\label{eq:StaticCaseSpinCurrent}
\frac{\gamma}{M_{s}} J \hat{\vect{x}} \cdot (\vect{M} \times \partial_{x} \vect{M}) + v_{s} M_{x} = v_{s}.
\end{equation}
Combining Eqs.~\eqref{eq:StaticCaseEnergy} and \eqref{eq:StaticCaseSpinCurrent} allows to calculate the static magnetization configuration for currents smaller than the critical current $j_{c}$, as discussed in Sec.~\ref{sec:CurrentBelowCriticalValue} below.

To determine $j_{c}$ we now evaluate Eqs.~\eqref{eq:StaticCaseEnergy} and \eqref{eq:StaticCaseSpinCurrent} at $x=0$.  Taking our boundary condition $\vect{M}(x=0) = \hat{\vect{z}}$ into account, we get:
\begin{equation*}
\frac{J}{2}(\partial_{x} \left.\vect{M})^{2}\right|_{x=0} - \lambda = 0
\quad \text{and} \quad
-\frac{\gamma}{M_{s}} J \left.\partial_{x} M_{y}\right|_{x=0} = v_{s}.
\end{equation*}
Furthermore, at $x=0$ we have $\partial_{x} \vect{M} \perp \hat{\vect{z}}$, and therefore $(\partial_{x} \vect{M})^{2} = (\partial_{x} M_{x})^{2} + (\partial_{x} M_{y})^{2}$.  Consequently we find $0 < J^{2} (\partial_{x} M_{x})^{2} = 2 J \lambda - (v_{s} M_{s}/\gamma)^2$.  Therefore, a static solution is possible if and only if the effective spin velocity $v_{s}$ is smaller than the critical spin velocity:
\begin{equation}
\label{eq:CriticalSpinVelocity}
v_{s}^{c} \equiv \frac{\gamma}{M_{s}} \sqrt{2 \lambda J},
\end{equation}
corresponding to the critical current $j_{c}$
\begin{equation}
\label{eq:CriticalCurrent}
j_{c} \equiv \frac{e M_{s}}{P \mu_{B}} v_{s}^{c} = \frac{e \gamma}{P \mu_{B}} \sqrt{2 \lambda J} = \frac{e \gamma \sigma}{P \mu_{B}} \sqrt{2 \lambda_{\mathrm{bulk}} J_{\mathrm{bulk}}}.
\end{equation}
In the last part of this equation we have explicitly written the critical current in terms of bulk parameters for comperison to the numerical results\footnote{In MicroMagnum, the convention for the exchange energy term is to use a sum over site indices rather than site links; therefore the exchange constant used in MicroMagnum is half the value of $J_{\mathrm{bulk}}$.}, as shown in Fig.~\ref{fig:fig3}.
Note that $j_c$ is smaller than the current above which a uniform ferromagnetic state becomes unstable, see below Sec.~\ref{sec:FerromagneticInstability}.

\section{Magnetization configuration for $j<j_{c}$}
\label{sec:CurrentBelowCriticalValue}

In this section, we find a general, static solution for currents below the critical current.  We show that such a magnetization configuration actually corresponds to a virtual DW that would be centered around a negative $x$ value when fictitiously extending the magnetic nano-wire also to negative $x$ values, as shown in Fig.~\ref{fig:fig2}.

For the magnetization described by a unit vector field, $\vect{M}^{2} = 1$, the derivative of the field is orthogonal to the field itself, i.e., $\vect{M} \perp \partial_{x} \vect{M}$.  Hence the vector field $\partial_{x} \vect{M}$ is two-dimensional and can be parametrized by two functions $\Lambda(x)$ and $\Gamma(x)$ as
\begin{equation}
\partial_{x} \vect{M} = (\hat{\vect{x}} \times \vect{M}) \Gamma(x) + [\vect{M} \times (\hat{\vect{x}} \times \vect{M})] \Lambda(x),
\end{equation}
whose $x$ component reduces to $\partial_{x} M_{x} = \Lambda(x) (1 - M_{x}^{2})$.  To simplify the notation we write $\Gamma$ and $\Lambda$ instead of $\Gamma(x)$ and $\Lambda(x)$, respectively.  The following calculation reveals that both $\Gamma$ and $\Lambda$ are finite for $0 < j < j_{c}$, implying that the current leads to a non-zero $y$ component of the magnetization texture, as can be seen from the simulation results plotted in Fig.~\ref{fig:fig1}.  In this parametrization, Eqs.~\eqref{eq:StaticCaseEnergy} and \eqref{eq:StaticCaseSpinCurrent} read:
\begin{gather}
\frac{J}{2} (1 - M_{x}^{2}) (\Lambda^{2} + \Gamma^{2}) = \lambda \Pi(M_{x}), \\
\Gamma (1 - M_{x}^{2}) = [M_{s}/(\gamma J)] v_{s} (1 - M_{x}).
\end{gather}
Eliminating $\Gamma$ in the above equations and using $\partial_{x} M_{x} = \Lambda (1 - M_{x}^{2})$ results in a partial differential equation,
\begin{multline}
\label{eq:PDE}
(\partial_{x} M_{x})^{2} = \biggl( \frac{M_{s}}{\gamma J} \biggr)^{2} \bigl[ (v_{s}^{c})^{2} \Pi(M_{x}) (1 - M_{x}^{2}) \\
- v_{s}^{2} (1 - M_{x})^{2} \bigr],
\end{multline}
which we can solve by separating variables:
\begin{multline}
\label{eq:PDESolution}
x = \frac{\gamma J}{M_{s}} \int_{0}^{M_{x}} dM_{x} \biggl[ (v_{s}^{c})^2 \Pi(M_{x}) (1 - M_{x}^{2}) \\
- v_{s}^{2} (1 - M_{x})^{2} \biggr]^{-1/2}.
\end{multline}
The above integral can be computed for any uniaxial magnetic anisotropy $\Pi (M_{x})$ and thus provide the full DW profile.

\subsection{Mapping the magnetic problem to an effective one dimensional model of a particle in a potential}
\label{sec:OneDimensionalModel}

\begin{figure}
\includegraphics[width=0.95\columnwidth]{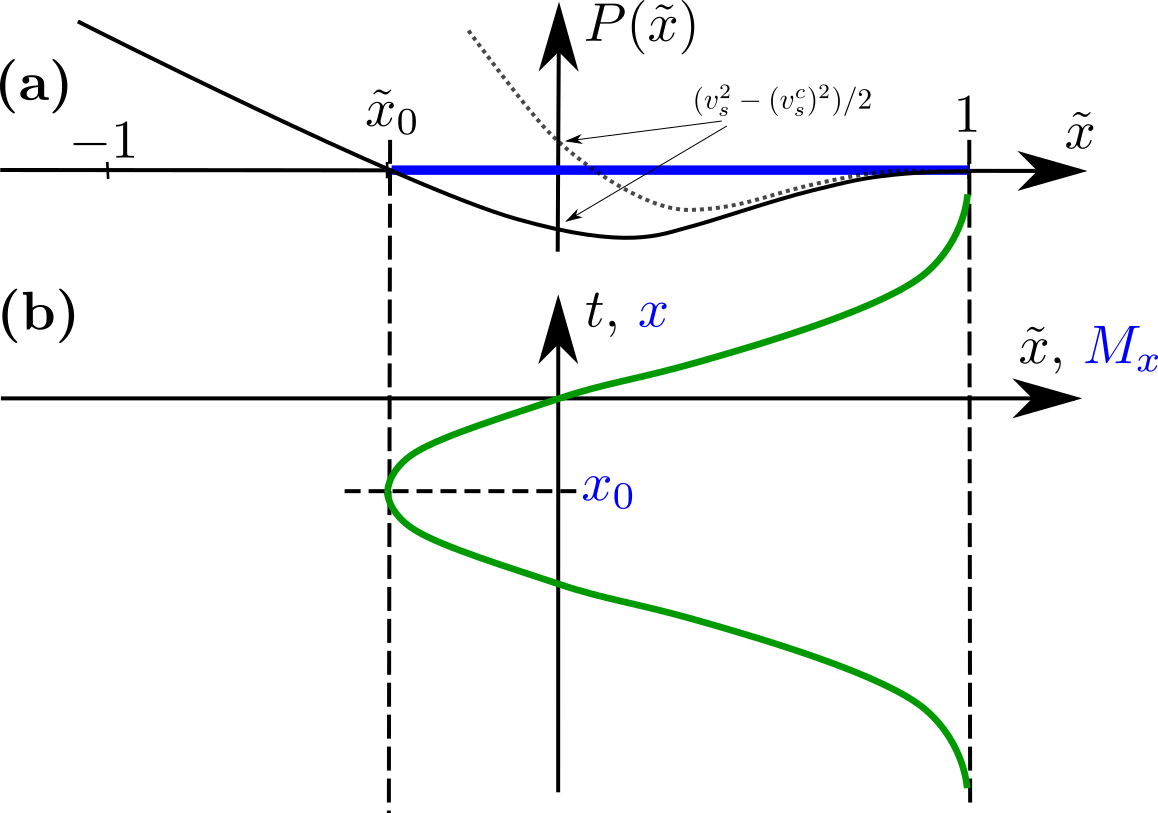}
\caption{(Color online) Upper graph (a): The solid black line is a sketch of a potential $P(\tilde{x})$ for $j < j_{c}$.  The allowed region for the particle in the potential with total energy zero is shown by the blue interval.  The dotted line shows the same function for $j > j_{c}$.  Lower graph (b): sketch of the function $\tilde{x}(t)$ or, translated into the language of the magnetic model, of $M_{x}(x)$.}
\label{fig:fig2}
\end{figure}

Let us take a closer look at Eq.~\eqref{eq:PDE} by changing the perspective and regarding the position $x$ along the wire as time $t$, and the magnetization $M_{x}$ in $x$ direction as the new spatial coordinate $\tilde{x}$.  Eq.~\eqref{eq:PDE} then transforms into
\begin{equation}
\frac{1}{2} \dot{\tilde{x}}^{2} \biggl( \frac{\gamma J}{M_{s}} \biggr)^{2} + P(\tilde{x}) = 0,
\end{equation}
which describes a one-dimensional fictitious particle of mass $(\gamma J/M_{s})^{2}$ and total energy $0$ moving in the potential, see upper part of the Fig.~\ref{fig:fig2}:
\begin{equation}
P(\tilde{x}) = \frac{v_{s}^{2}}{2} (1 - \tilde{x})^{2} - \frac{(v_{s}^{c})^{2}}{2} \Pi(\tilde{x}) (1 - \tilde{x}^{2}).
\end{equation}
At the end of the wire, the magnetization direction aligns along the uniaxial anisotropy direction, $M_{x}(x \to \infty) = 1$, which translates into $\tilde{x}(t \to \infty) = 1$.  This implies:
\begin{equation*}
P(\tilde{x} \sim 1) \approx \biggl[ \frac{v_{s}^{2}}{2} + (v_{s}^{c})^{2} \Pi'(1) \biggr] (1 - \tilde{x})^{2}.
\end{equation*}
This gives us an important physically relevant insight into the stability of the solution. The function $P(\tilde{x})$ has the following properties: $P(0) = [v_{s}^{2} - (v_{s}^{c})^{2}] / 2$, $P(1) = 0$, and $P'(1)=0$. $P''(1)$ is negative for $v_s^2 <-({v_{s}^c})^2  \Pi'(1)$
(according to our definition $\Pi'(1)$ is negative) and becomes positive for  $v_s^2 >-({v_{s}^c})^2  \Pi'(1)$.
 However at current
$v_{s}^{2} > (v_{s}^{c})^{2}$, $P(0)$ becomes positive. Note, that for any reasonable physical system $-\Pi'(1) >1$, so there are a range of currents $(v_{s}^{c})^{2}<v_{s}^{2}< -({v_{s}^c})^2  \Pi'(1)$ at which $\tilde{x}$ (i.e. $M_x$) is never zero during the motion, see Fig.~\ref{fig:fig2}, so that in the magnetic system the boundary condition at the pinned center cannot be satisfied, while the uniform ferromagnetic state is still stable.

Our magnetization problem now corresponds to a particle that at time zero is at the origin, $\tilde{x}(t=0) = 0$, and that approaches unity, $\tilde{x}(t \to \infty) = 1$.  Since energy is conserved, this particle might have the following history:  the motion of the particle starts at $t = -\infty$ at $\tilde{x} = 1$ with an infinitesimal negative initial velocity.  Then the particle moves to the left and at some time $t_{0}$ reaches the turning point $\tilde{x}_{0}$ defined by $P(\tilde{x}_{0}) = 0$ where it switches the direction of the motion, see the lower part of the Fig.~\ref{fig:fig2}.  Note that only for $v_{s} < v_{s}^{c}$ the turning point is at negative $\tilde{x}$ values, $\tilde{x}_{0} < 0$, so that the particle crosses $\tilde{x}=0$ ($M_x=0$) twice. The origin of time $t=0$ ($x=0$) then is defined as a time when the fictitious particle crosses   $\tilde{x}=0$ ($M_x=0$) for the second time, see lower graph of Fig.~\ref{fig:fig2}.

Such a solution is, of course, symmetric with respect to the time $t_{0}$ where the particle reverses its direction:
\begin{equation}
t_{0} = - \frac{\gamma J}{M_{s} \sqrt{2}} \int_{\tilde{x}_{0}}^{0} \frac{d\tilde{x}}{\sqrt{-P(\tilde{x})}}.
\end{equation}
For a current close to $j_{c}$ the turning point is small, $\tilde{x}_{0} \lesssim 0$, so that we can approximate the potential around $\tilde{x}= 0$ as $P(\tilde{x}) \approx v_{s}^{c} (v_{s} - v_{s}^{c}) - (v_{s}^{c})^{2} \tilde{x}$ from which we can determine $\tilde{x}_{0} = (v_{s} - v_{s}^{c})/v_{s}^{c}$.  For $v_{s} \lesssim v_{s}^{c}$ we thus obtain
\begin{equation*}
\begin{split}
t_{0} &= - \frac{\gamma J}{M_{s} \sqrt{2}}\int_{(v_{s} - v_{s}^{c})/v_{s}^{c}}^{0} \frac{dx}{\sqrt{v_{s}^{c} (v_{s}^{c} - v_s) + (v_{s}^{c})^{2} x}} \\
&= - \frac{\gamma J \sqrt{2}}{M_{s} v_{s}^{c}} \sqrt{\frac{v_{s}^{c} - v_s}{v_{s}^{c}}}.
\end{split}
\end{equation*}

\subsection{Interpretation as a part of a domain wall}
\label{sec:DomainWallInterpretation}

Let us now translate the problem of the particle in a potential well back to our magnetic model, i.e., time translates back into the spatial coordinate of the magnetic wire, and the position $\tilde{x}$ of the particle corresponds to the magnetic component $M_{x}$ along the uniaxial anisotropy direction.  Extending the problem of the particle to negative times corresponds to fictitiously extending the semi-infinite wire also to negative spatial coordinates.  The plot of position of the particle versus time, shown in Fig.~\ref{fig:fig2}, which is symmetric w.r.t.\ the maximum defined by the turning time $t_{0}$ and coordinate $\tilde{x}_{0}$ then corresponds to a plot of magnetic component $M_{x}$ versus spatial coordinate of the wire.  In other words, it shows the profile of the $x$ component of the magnetization along the fictitiously extended wire, displaying a DW centered around the coordinate $x_{0} < 0$ in the unphysical region with
\begin{equation}
\label{eq:DomainWallPosition}
x_{0} = - \frac{\gamma J \sqrt{2}}{M_{s} v_{s}^{c}} \sqrt{\frac{v_{s}^{c} - v_s}{v_{s}^{c}}} =  - \frac{\gamma J e \sqrt{2}}{P \mu_{B} j_{c}} \sqrt{\frac{j_{c} - j}{j_{c}}}.
\end{equation}
We can also estimate the current-dependent width $\Delta_{j}$ of the fictitious DW by $\Delta_{j}^{-2} \approx \partial_{x}^{2} M_{x}(x=0)$.  Using Eq.~\eqref{eq:PDESolution} we obtain
\begin{equation}
\label{eq:DomainWallWidth}
\Delta_{j} \approx \frac{\gamma J}{M_{s}} \biggl[ v_{s}^2 + \frac{(v_{s}^{c})^2}{2} \Pi'(1) \biggr]^{-1/2}.
\end{equation}
Furthermore, the fictitious DW is not planar. It is twisted around the axis defined by the uniaxial anisotropy. The characteristic, current strength dependent pitch is
\begin{equation}
\Delta_{\Gamma_{j}}\approx (\partial_x M_{y}(x=0))^{-1} = \frac{\gamma J}{M_s v_s}.
\end{equation}
Using this length scale we can rewrite Eq. \eqref{eq:DomainWallPosition} as
\begin{equation}
x_{0} = - \sqrt{2}\, \Delta_{\Gamma_{j_c}} \sqrt{\frac{v_{s}^{c} - v_s}{v_{s}^{c}}},
\end{equation}
which shows that the characteristic twisting length scale sets the length scale for process of DW production.

Increasing the current towards the critical value moves the fictitious center of the DW towards the pinned end of the wire at $x=0$.  This provides an intuitive picture as to why for currents above the critical current the DW will ``twist off'' and move along the wire.  But before we discuss the dynamic solution we first address the ferromagnetic instability of the system and check that this occurs for even larger currents than $j_{c}$.

\subsection{The ferromagnetic instability}
\label{sec:FerromagneticInstability}

The analysis of the potential close to $1$, where we obtained the condition $v_{s}^{2}/2 +(v_{s}^{c})^{2} \Pi'(1) < 0$, yields that above a certain effective critical spin velocity, i.e., above a critical current $j_{c*}$ even the ferromagnetic solution becomes unstable.  To be precise, the boundary condition at infinity, $M_{x}(x \to \infty) = 1$, originating in the uniaxial anisotropy, can only be satisfied up to the current $j_{c*} = j_{c} \sqrt{-2 \Pi'(1)}$ or $v_{s}^{c*} = v_{s}^{c} \sqrt{-2 \Pi'(1)}$, respectively.  For the standard form of the uniaxial anisotropy, $\Pi(M_{x}) = 1 - M_{x}^{2}$, this means that the current above which the ferromagnetic solution is unstable, is twice as large as the critical current above which domain walls are created, $j_{c*} = 2 j_{c}$.  Note that this stability condition also implies a condition on the form of the uniaxial anisotropy, as only for $-2 \Pi'(1) > 1$ it is $j_{c*}> j_{c}$ and domain walls are created before the ferromagnetic solution breaks down.  For the idea of how the ferromagnetic istability is reached it is instructive to look again at the equation of the DW width, cf.\ Eq.~\eqref{eq:DomainWallWidth}.  With our definition of $j_{c*}$ or $v_{s}^{c*}$, respectively, we can write
\begin{equation}
\label{eq:DomainWallWidth2}
\Delta_{j} \approx \frac{\gamma J}{M_{s}} \bigl[ v_{s}^{2} - (v_{s}^{c*})^{2} \bigr]^{-1/2},
\end{equation}
so the width of the DW diverges at the current $j = j_{c*}$.

\begin{figure}
\includegraphics[width=0.95\columnwidth]{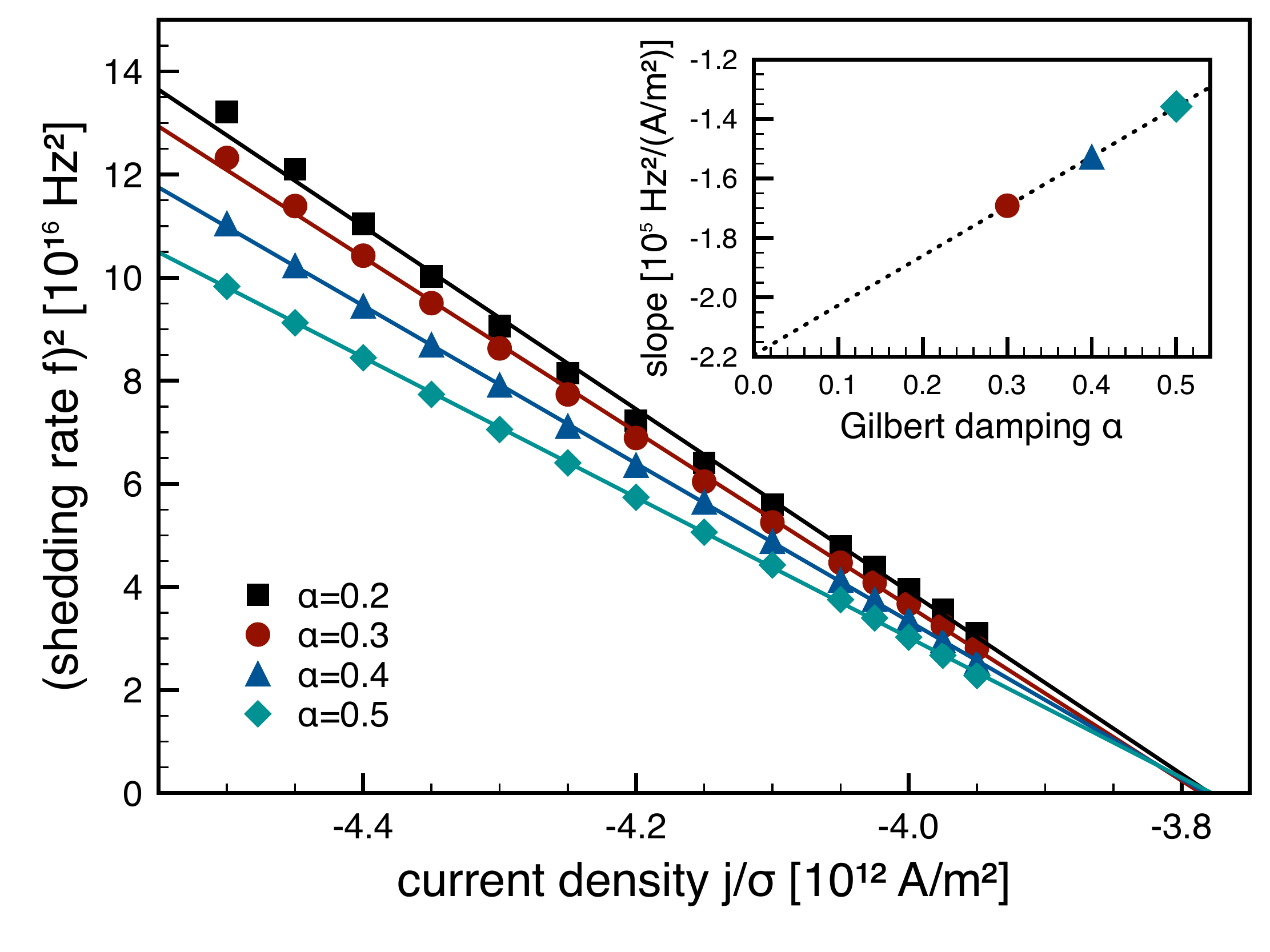}
\caption{(Color online) Numeric results based on MicroMagnum\cite{MicroMagnum} for the square of the frequency of texture formation as a function of current for different $\alpha$.  For these simulations we used\cite{Note3} $\lambda_{\mathrm{bulk}} = \unit{10^{4}}{\joule\per\meter\cubed}$, $J_{\mathrm{bulk}} = \unit{1.6 \cdot 10^{6}}{\joule\per\meter}$, and we have fixed the magnetization at the start of the wire by a local magnetic field.  For these values, numerically we obtain for the critical current density $j_{c}^{\mathrm{num}}/\sigma \approx \unit{-3.78 \cdot 10^{12}}{\ampere\per\meter\squared}$ which agrees within four percent with the analytical result of $j_{c} \approx \unit{-3.92 \cdot 10^{12}}{\ampere\per\meter\squared}$.  The small discrepancy is probably due to the method of how we fixed the spin in the numerics.  Inset: slope of the main figure vs damping constant $\alpha$.  At $\alpha = 0$ we can compare to the analytical result of $\unit{-2.5 \cdot 10^{5}}{\ampere\per(\second\usk\meter\squared)}$ and see that we get the right order.}
\label{fig:fig3}
\end{figure}

\section{Dynamics of DW creation}
\label{sec:DynamicCreationOfDWs}

In this part we consider the magnetization dynamics for currents (just) above $j_{c}$, where a static solution no longer exists, see, e.g., Eq.~\eqref{eq:DomainWallPosition}, and therefore we must look for a time-dependent solution.  As the current increases towards $j_c$ the virtual DW approaches the start of the magnetic wire from the left, thus the major effect for currents just above $j_c$ will be a moving DW, so a time-dependent coordinate of the center of the DW.  Therefore, we look for the solution of the form
\begin{equation}
\label{eq:DynamicCaseAnsatz}
\vect{M}(x, t) = \vect{M}_{0}(x - x_{0}(t); v_{s_{x_{0}(t)}}) + \vect{s},
\end{equation}
where $\vect{M}_{0}(x - x_{0}; v_{s_{x_{0}}})$ is the static solution with effective spin velocity $v_{s_{x_{0}}}$ that solves Eq.~\eqref{eq:DomainWallPosition} for a given fictitious DW center $x_{0}$, and the vector field $\vect{s}$ is a small perturbation, $|\vect{s}| \ll 1$.  As $\vect{M}(-x_{0}(t); v_{s_{x_{0}}}) = \hat{\vect{z}}$ and $\vect{M}(x - x_{0}(t); v_{s_{x_{0}}}) \to \hat{\vect{x}}$ for large $x$, the vector field $\vect{s}$ has to vanish at the start and end of the semi-infinite wire, $\vect{s}(x = 0, t) = 0$ and $\vect{s}(x \to \infty, t) \to 0$.

To test our ansatz of Eq.~\eqref{eq:DynamicCaseAnsatz} let us plug it in into the simplified LLG equation, Eq.~\eqref{eq:SimplifiedLLG}.  For this we need to calculate first the time derivative of Eq.~\eqref{eq:DynamicCaseAnsatz}:
\begin{equation}
\partial_{t} \vect{M} = -\dot{x}_{0} \partial_{x} \vect{M}_{0} + \frac{\partial \vect{M}_{0}}{\partial v_{s_{x_{0}}}} \frac{\partial v_{s_{x_{0}}}}{\partial x_{0}} \dot{x}_{0} +\frac{\partial \vect{s}}{\partial t}.
\end{equation}
For currents just above the critical current, the second term on the RHS is small compared to the other ones, as $x_0$ is small and this term is higher order in $x_0$, because  $\frac{\partial v_{s_{x_{0}}}}{\partial x_{0}}\sim x_{0}$.  Now plugging Eq.~\eqref{eq:DynamicCaseAnsatz} into Eq.~\eqref{eq:SimplifiedLLG} and linearizing Eq.~\eqref{eq:SimplifiedLLG} in the vector field $\vect{s}$ we obtain an inhomogeneous linear equation that needs to be solved for the vector field $\vect{s}(x)$  satisfying above boundary equations:
\begin{multline}
- \frac{\partial \vect{s}}{\partial t} + \vect{s} \times \frac{\delta F}{\delta \vect{M}_{0}} + \vect{M}_{0} \times \frac{\delta^{2} F}{\delta \vect{M}_{0}^{2}} \vect{s} - v_{s} \frac{\partial \vect{s}}{\partial x} \\
= (v_{s} - v_{s_{x_{0}}} - \dot{x}_{0}) \partial_{x} \vect{M}_{0}.
\end{multline}
The trivial solution of above equation, $\vect{s} \equiv \vect{0}$, that is in agreement with the vector field $\vect{s}$ vanishing at the start and end of the wire, exists if the applied current obeys the following equation:
\begin{equation}
\dot{x}_{0} = v_{s} - v_{s_{x_{0}}} = v_{s} - v_{s}^{c} + \frac{M_{s}^{2} (v_{s}^{c})^3}{2 \gamma^{2} J^{2}} x_{0}^{2},
\end{equation}
where we obtained $v_{s_{x_{0}}}$ via Eq.~\eqref{eq:DomainWallPosition}.  Thus
\begin{equation}
t = \int dx_{0} \biggl[ v_{s} - v_{s}^{c} + \frac{M_{s}^{2} (v_{s}^{c})^{3}}{2 \gamma^{2} J^{2}} x_{0}^{2} \biggr]^{-1}.
\end{equation}
The main contribution to  the integral comes from $x_{0}^{2} \sim 2 \gamma^{2} J^{2} (v_{s} - v_{s}^{c})/[M_{s}^{2} (v_{s}^{c})^3] \to 0$, and outside of this region the integral converges very quickly. So in finding the period $T$ it is justified to extend the integration to infinity, even though the initial equation is correct only for small $x_{0}$ --- the error will be exponentially small as $(v_{s} - v_{s}^{c}) \to 0$.  Finally. we obtain our central analytical result for $j > j_{c}$:
\begin{equation}
\label{eq:PeriodDynamicDomainWalls}
\begin{split}
T &= \int_{-\infty}^{\infty} dx_{0} \biggl[ v_{s} - v_{s}^{c} + \frac{M_{s}^{2} (v_{s}^{c})^{3}}{2 \gamma^{2} J^{2}} x_{0}^{2} \biggr]^{-1} \\
&= \frac{\sqrt{2} \pi \gamma J}{M_{s} (v_{s}^{c})^2} \sqrt{\frac{v_{s}^{c}}{v_{s} - v_{s}^{c}}} \\
&= \frac{\sqrt{2} \pi e^{2} J M_{s} \gamma}{j_{c}^{2} P^{2} \mu_{B}^{2}} \sqrt{\frac{j_{c}}{j - j_{c}}}.
\end{split}
\end{equation}
In particular it shows that for currents just above the critical current $j_{c}$ new DWs are injected from the pinned start of the wire periodically with a frequency $f = T^{-1} \sim \sqrt{j - j_{c}}$.

We confirm our analytical result also within simulations, see Fig.~\ref{fig:fig3}, where $f^{2}$ as a function of the applied current density $j$ is plotted.  Note that the obtained value for the critical current in the numerics is independent of the damping parameter $\alpha$,
and is of the right value $j_{c}^{\mathrm{num}}/\sigma \approx \unit{-3.78 \cdot 10^{12}}{\ampere\per\meter\squared}$ compared to analytics
$j_{c} \approx \unit{-3.92 \cdot 10^{12}}{\ampere\per\meter\squared}$.  However the slope does depend on $\alpha$, as one expects that larger damping slows down the dynamics and less DWs will be produced.  The DW production dynamics is difficult to simulate at small values of $\alpha$ as it takes a very long time for the system to reach a steady state at currents close to $j_{c}$.  From our numerical data we can infer that the slope value for $\alpha = 0$ is of the order of $\unit{-2.2 \cdot 10^{5}}{\ampere\per(\second\usk\meter\squared)}$ which is in the right order of the analytical value of $\unit{-2.5 \cdot 10^{5}}{\ampere\per(\second\usk\meter\squared)}$.

\section{Discussion}
\label{sec:Discussion}

The essential parts of the analytic calculations where shown in the specific geometry, as shown in Fig.~\ref{fig:fig1}.  However, all of our results are independent of the specific underlying model and the peculiar directions.  The only requirements are exchange interactions and uniaxial anisotropy along a different direction than the pinned magnetization at the one end of the wire. In particular we would like to stress, that the direction of the uniaxial anisotropy does not matter. A uniaxial anisotropy in a direction perpendicular to the wire would also lead to periodic domain wall formations with the same underlying mechanisms as long as the pinned magnetization in the wire is at a finite angle to the uniaxial anisotropy direction.  Furthermore, it is of course not essential that the wire is semi-infinite (it just needs to be "long enough"), neither is it essential that the magnetization is fixed at the \emph{pinned end} of the wire. The same results are valid for a process of a current driven DW crossing a very strong pinning center \emph{in} a nano-wire. The pinning center pins a DW to itself, but another DW will cross it if the current is above the critical current $j_{c}$.

In our simulations we have used the exchange parameter of Permalloy $J_{\mathrm{bulk}} = \unit{1.6 \cdot 10^{6}}{\joule\per\meter}$, and $\lambda_{\mathrm{bulk}} = \unit{10^{4}}{\joule\per\meter\cubed}$ which is reasonable for a magnetic nanowire, see Fig.~\ref{fig:fig3}.  With these parameters we obtained about $4 \cdot \unit{10^{8}}{\ampere\per c\meter\squared}$ for the critical current density.

To reduce the problems with Ohmic heating in experiments one might consider to apply current pulses. In order to inject a single DW the current must be nonzero only for the time given by the period $T$.  The total amount of the Ohmic heat per created DW is about $j^{2}T\sim j^{2}/\sqrt{(j/j_{c})^{2} - 1}$.  So the amount of heat produced per DW is minimal at $j = \sqrt{2}\ j_{c}$ (which is still smaller than $j_{c*}$ for the standard anisotropy term).  The total amount of heat produced in the process is $\sim 2 j_{c}^{2}$.

Our work here has an important perspective from dynamic system theory.
The vast majority of self-oscillations (limit cycles) induced by spin-transfer in magnetic systems correspond to a \emph{dynamic} loss of stability, i.e., generically to a pair of conjugated eigenfrequencies crossing the real axis from positive to negative damping, resulting into an Andronov-Hopf bifurcation. This is not the case here. As a matter of fact, Andronov-Hopf bifurcations always manifest themselves with a vanishing oscillation amplitude and a finite period at the bifurcation point, i.e., at the critical current value. Here, in strong contrast, we derive analytically and observe numerically a finite, in fact saturated, amplitude of oscillation and a diverging period at the critical current value. These unique characteristics are strong indications of a saddle-node homoclinic bifurcation, consistent with the established \emph{static} loss of stability, and put the considered system in a class of itself as far as spin transfer induced oscillations are concerned.

\section{Conclusion}
\label{sec:Conclusion}

To conclude, we have considered a ferromagnetic nano-wire with a strong pinning center.  We have shown numerically and analytically that
as one increases the current the magnetic texture at the pinning center stretches and twists until above a certain critical current a domain wall ``twist off'' from the impurity and travels along the nano-wire, Fig.~\ref{fig:fig1}.  For currents above the critical current this process happens periodically.

The period at which this happens is given by a universal exponent, $T\sim (j - j_{c})^{-1/2}$.  We emphasize, that this process is very general, and is independent of microscopic details.  It occurs already in very simple systems exhibiting only exchange and uniaxial anisotropy interactions.

A key message from our result is that the process of domain wall injection by currents requires neither any ``twisting'' terms in the model, such as DMI, or dipole-dipole interactions, nor an assisting magnetic field.  We expect that the DMI will not change the results considerably, but will lower the critical current.  The effects of the dipole-dipole interaction are harder to estimate, as in particular their exact form will depend on the realization of the geometry.

We predict that current-induced periodic DW production will be observable in simple Permalloy nano-wires.

\begin{acknowledgments}
Ar.~A.\ is grateful to O.~A.\ Tretiakov for numerous discussions and to Bin Yang who participated at the start of the project.  Ar.~A.\ is also very grateful for the warm hospitality of the supporting staff of the INSPIRE group at Johannes Gutenberg-Universit{\"a}t, Mainz, Germany. J.~S.\ and Ar.~A.\ acknowledge the support of the Alexander von Humboldt Foundation.  K.~E.-S.\ acknowledges funding by the SFB TRR 173 (``Spin+X'').

Micromagnetic simulations were performed based on MicroMagnum\cite{MicroMagnum}, and M.~S.\ and K.~E.-S.\ are grateful for many helpful discussions with K.~Litzius.
\end{acknowledgments}

\bibliography{references-mendeley,references}

\end{document}